\documentclass[twocolumn,showpacs,preprintnumbers,amsmath,amssymb,floatfix]{revtex4}
\usepackage[dvips]{graphicx}

\usepackage{dcolumn}
\usepackage{bm}
\usepackage{color}

\begin{document}
\title{Statistical Reconstruction of Qutrits}

\author{Yu.I.Bogdanov}
 \affiliation{Russian Control System Agency, ``Angstrem'', Moscow 124460 Russia.}

\author{M.V.Chekhova,
L.A.Krivitsky, S.P.Kulik, A.N.Penin, A.A.Zhukov}
 \email{postmast@qopt.phys.msu.su}
\affiliation{Department of Physics, Moscow M.V. Lomonosov State
University, 119992 Moscow, Russia.}

\author{L.C.Kwek}
 \email{lckwek@nie.edu.sg}
 \affiliation{National Institute of Education Nanyang
Technological University, 637616 Singapore.}

\author{C.H.Oh, M.K.Tey}
 \email{phyohch@nus.edu.sg, phyteymk@nus.edu.sg}
 \affiliation{Department of Physics, Faculty of Science, National University of Singapore,
117542 Singapore.}

\begin{abstract}
We discuss a procedure of measurement followed by the reproduction
of the quantum state of a three-level optical system - a
frequency- and spatially degenerate two-photon field. The method
of statistical estimation of the quantum state based on solving
the likelihood equation and analyzing the statistical properties
of the obtained estimates is developed. Using the root approach of
estimating quantum states, the initial two-photon state vector is
reproduced from the measured fourth moments in the field . The
developed approach applied to quantum states reconstruction is
based on the amplitudes of mutually complementary processes.
Classical algorithm of statistical estimation based on the Fisher
information matrix is generalized to the case of quantum systems
obeying Bohr's complementarity principle. It has been
experimentally proved that biphoton-qutrit states can be
reconstructed with the fidelity of 0.995-0.999 and higher.
\end{abstract}

\maketitle

\section{Introduction}

The ability of measuring quantum states is of fundamental interest
because it provides a powerful tool for the analysis of basic
concepts of quantum theory, such as the fundamentally statistical
nature of its predictions, the superposition principle, Bohr's
complementarity principle, etc. To measure quantum state one needs
to perform some projective measurements on the state and then to
apply some computation procedure to the data. The first step is a
genuine measurement consisting of a set of operations on the
representatives of a quantum statistical (pure or mixed) ensemble.
As a result of such operation an experimentalist acquires a set of
frequencies at which particular events occur. In the second step a
mathematical procedure is applied to the statistical data obtained
in the previous step to reconstruct the quantum state. The present
paper is devoted to the state reconstruction for the optical
three-level systems. The object under study is the polarization
state of a frequency- and spatially degenerate biphoton
field~\cite{bur}.

The necessity of the adequate measurement of the states of such
systems is caused not only by fundamental interest but also by
some applications. For example, it has been shown that the
security of the key distribution in quantum cryptography is
associated with the dimensionality of the Hilbert space for the
states in use ~\cite{bech}. From this point of view certain hopes
are pinned on the three-level systems or
qutrits~\cite{bech1,cerf,bruk} rather than qubits.

We should mention that there are other implementations of
three-level optical systems. The most familiar ones deal with
three-arm interferometers~\cite{thew} and lower-order transverse
spatial modes of optical field, realized with
holograms~\cite{vaziri,molina,lang}. Polarization--entangled
four-photon fields, which are equivalent to two entangled spin-1
particles were studied in~\cite{how}.

The paper is organized as follows. In Sec.II we discuss the main
properties of qutrits based on the polarization state of biphoton
field. We focus on their preparation, visual representation on
Poincar\'{e} sphere, unitary transformation by phase plates. Then
we consider the coherence matrix, which characterizes completely
the properties of biphotons-qutrits in the fourth field moments.
Sec.III is devoted to the methods of biphotons-qutrits
measurement, in particular, we introduce two quantum tomography
protocols and discuss in detail their experimental implementation.
We conclude this part with the analysis of statistical
reconstruction for qutrits from the outcomes of mutually
complementary measurements. Sec. IV deals with the methods of
quantum state reconstruction. Namely we consider the least-squares
and maximum-likelihood methods and apply these tools to analysis
the data obtained in quantum tomography. In Appendix we explore
the problem of statistical fluctuations of the state vector which
is important for the estimation and control of precision and
stability of quantum information.

\section{Qutrits based on biphotons}

\subsection{Preparation}

Biphoton field is a coherent mixture of two-photon Fock states and
the vacuum state~\cite{klbook}:

\begin{equation} \Psi = | {vac} \rangle + \frac{1}{2}\sum\limits_{\vec{k_s}\vec{k_i}} {F_{\vec{k_s},\vec{k_i} }
| {1_{\vec{k_s} } ,1_{\vec{k_i} } } \rangle } ,\end{equation}
where $| {1_{\vec{k_s} } ,1_{\vec{k_i} } } \rangle $ denotes the
state with one (signal) photon in the mode $\vec{k_{s}}$ and one
(idler) photon in the mode $\vec{k_{i}}$. The coefficient
$F_{\vec{k_s} ,\vec{k_i} } $ is called the biphoton amplitude
~\cite{kllaser}, because its squared modulus gives a probability
to register two photons in modes $\vec{k_{s}}$ and $\vec{k_{i}}$.

Let us consider the collinear and frequency degenerate regime, for
which $\vec {k}_s \approx \vec {k}_i $, $\omega _s \approx \omega
_i $ and $\omega _s + \omega _i = \omega _p $, where $\omega _p $
is the laser pump frequency. We further restrict our discussion to
biphotons that are indistinguishable in terms of spatial,
spectral, or temporal parameters. From the point of view of
polarization there are three natural states of biphotons, namely,
$\Psi _1 = |{2,0} \rangle $, $\Psi _2 = | {1,1} \rangle $, and
$\Psi _3 = | {0,2} \rangle $. Here the notation $| {2,0} \rangle
\equiv | {2_H ,0_V } \rangle $, for example, indicates that there
are two photons in the horizontal $(H)$ polarization mode, while
no photons are present in the orthogonal vertical $(V)$ mode.
These basic states can be generated using type-I (for $\Psi _1 $
and $\Psi _3$ ) and type-II (for $\Psi _2 $) phase matching. Since
only two-photon Fock states are considered, for the state $| {m,n}
\rangle $ the condition $m + n = 2$ must be satisfied.

Any arbitrary pure polarization state of biphoton field can be expressed in
terms of three complex amplitudes $c_1 ,c_2 ,$ and $c_3 $:

\begin{equation} | c \rangle = c_1 | {2,0} \rangle + c_2 | {1,1} \rangle + c_3 |
{0,2} \rangle , \end{equation}where $c_j = | {c_j } |\exp \{
{i\varphi _j } \}$, $\sum\limits_{j = 1}^3 {| {c_j } |^2 = 1} $.
The vector $| c \rangle = ( {c_1 ,c_2 ,c_3 } )$ represents a
three-state state or qutrit.

There is an important note concerning the state-vector (2). In
principle, one can write the complete polarization state in the
form

\begin{equation}| c \rangle = c_1 | {2_H ,0_V } \rangle + c_2 | {1_V ,1_H }
\rangle + {c}'_2 | {1_H ,1_V } \rangle + c_3 | {0_H ,2_V } \rangle
, \end{equation} where the terms $| {1_H ,1_V } \rangle $ and $|
{1_V ,1_H } \rangle $ might be distinguishable somehow, for
example, if the photon with vertical polarization comes first with
respect to the photon with horizontal polarization. However we
consider particular two-mode polarization state so photons differ
in polarization only and there are no other parameters responsible
for their distinguishability.

In general, to generate an arbitrary qutrit state one needs to put
three nonlinear crystals separated in space into a common pump and
superpose the biphoton fields generated by the three crystals
coherently or incoherently (Fig.1).
\begin{figure}
\includegraphics[width=0.3\textwidth]{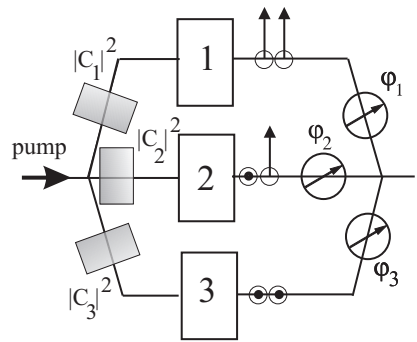}
\caption{Preparation of an arbitrary qutrit based on biphotons, in
principle. Three nonlinear crystals placed in the common pump
generate biphotons with type-I (1, 3) and type-II (2)
phase-matching. Three attenuators ($| {c_1 } |^2,| {c_1 } |^2,|
{c_3 } |^2)$ and three phase-shifters ($\varphi _1 ,\varphi _2
,\varphi _3 )$ allow one to control three complex amplitudes
$c_{1}$, $c_{2}$ and $c_{3}$.}
\end{figure}

\subsection{Representation of qutrits using the Poincar\'{e}
sphere}

Sometimes it is very convenient to use visual representation of
the state. For example a single-photon pure polarization state
(qubit) may be mapped onto the Poincar\'{e} sphere (3-dimensional
Euclidian sphere). A (pure) qubit state is determined by polar and
azimuthal angles $(\vartheta ,\phi )$ in spherical coordinates.
Any unitary polarization transformation of the qubit is
represented by the corresponding rotation of the sphere. Thus, in
order to learn the final transformed state one just has to apply
the rotation operation using certain rules.

It would be helpful to use same visual representation of a qutrit
using the Poincar\'{e} sphere. Although the generalization of the
Poincar\'{e} sphere for qutrits has been discussed earlier
~\cite{arvid} we suggest an alternative approach, which allows us
to manipulate with qutrits in natural 3-D space rather than in
sophisticated 8-D space. Let us map the polarization state of a
biphoton into a pair of points on the sphere (but this is not the
two-qubit case since the states$\left| {H,V} \right\rangle $ and
$\left| {V,H} \right\rangle $ are indistinguishable). In this
representation each photon forming the biphoton is plotted as a
single point on the Poincare sphere, so the qutrit state vector is
represented by

\begin{equation}| c \rangle = \frac{\left[a_s^\dag ( {\vartheta ,\phi
} )a_i^\dag ( {\vartheta',\phi'})+a_i^\dag ( {\vartheta ,\phi }
)a_s^\dag ( {\vartheta',\phi'})\right]|vac\rangle}{\left|\left|
\left[a_s^\dag ( {\vartheta ,\phi } )a_i^\dag (
{\vartheta',\phi'})+a_i^\dag ( {\vartheta ,\phi } )a_s^\dag (
{\vartheta',\phi'})\right] |vac\rangle
\right|\right|}\end{equation} where $a^\dag ( \vartheta _i ,\phi
_i )$ and $a^\dag (\vartheta _s ,\phi _s )$ are the creation
operators in idler and signal polarization modes and
$a^{\dagger}(\vartheta_m, \phi_m)
=\cos(\vartheta_m/2)a^{\dagger}+{\rm
e}^{i\phi_m}\sin(\vartheta_m/2)b^{\dagger}, \emph{m=i,s}$. Note
that operators $a^\dag \equiv a_H^\dag ,b^\dag \equiv a_V^\dag $
are creation operators for $H-$ and $V-$polarized photons.

It is well-known that the number of real parameters characterizing
a quantum state is determined by the dimension of the Hilbert
space $(s)$. For a pure state,
\begin{subequations}
\begin{equation} N_{pure} = 2s-2,
\end{equation} and for mixed states,
\begin{equation}N_{mixed} = s^{2}-1.\end{equation}\end{subequations}
 According to (5a,b), four real parameters
determine completely the pure state of a qutrit, so in the
Poincar\'{e} sphere representation these parameters are simply the
four spherical angles $(\vartheta _i ,\phi _i ;\vartheta _s ,\phi
_s)$. The links between the angles $( \vartheta _i ,\phi _i
;\vartheta _s ,\phi _s
 )$ and the amplitudes $c_j = |c_j|\exp{ i\varphi_j }$ are derived in ~\cite{burjetp}. As an example
three basic states $\Psi _1 = | {2,0} \rangle $, $\Psi _2 = |
{1,1} \rangle $, and $\Psi _3 = | {0,2} \rangle $ are shown in
Fig.2. It can be shown that the polarization degree of a qutrit $P
= \sqrt {|c_1|^2 - |c_3|^2 + 2| {c_1^\ast c_2 + c_2^\ast c_3 }
|^2} $ ~\cite{bur} has a clear geometrical meaning: it is defined
by the angle $\beta $ between the pair of points on the Poincare
sphere as seen from its center:

\begin{figure}
\includegraphics[width=0.3\textwidth]{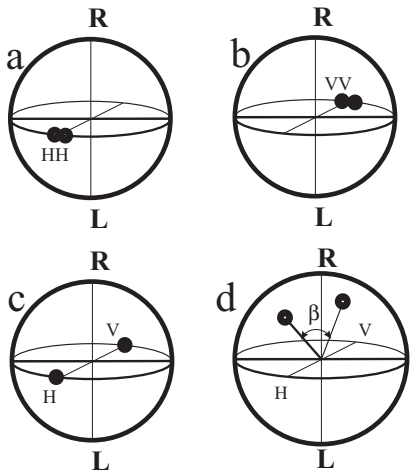}
\caption{Representation of a qutrit using the Poincar\'{e} sphere.
Figures \textit{a, b, c} show three basic states forming
superposition (2). Figure \textit{d} represents the state of an
arbitrary qutrit.}
\end{figure}

\begin{equation}P = \frac{2\cos ( \beta/2 )}{1 + \cos
^2( {\beta/2} )}.\end{equation} For the states $\Psi _1 $ and
$\Psi _3 $ the polarization degree takes values $P_{1,3} = 1$,
since two points coincide on the sphere and $\beta = 0$. For the
second state, $\Psi _2 $, two points are positioned at the
opposite sides of the sphere, that is why $\beta /2 = \pi/2$ and
$P_2 = 0$.

\subsection{Transformation}

Experimentally a unitary transformation of the polarization state
(2) can be achieved by placing any retardation plates, rotators
etc. into the biphoton beam. The action of such elements on the
state (2) is described by the matrix~\cite{burpra}:

\begin{equation}G = \left( {{\begin{array}{*{20}c}
 {t^2} \hfill & {\sqrt 2 tr} \hfill & {r^2} \hfill \\
 { - \sqrt 2 tr^ * } \hfill & {\vert t\vert ^2 - \vert r\vert ^2} \hfill &
{\sqrt 2 t^ * r} \hfill \\
 {r^{ * 2}} \hfill & {-\sqrt 2 t^ * r^ * } \hfill & {t^{ * 2}} \hfill \\
\end{array} }} \right),
\end{equation} where
\begin{equation} t = \cos \delta + i\sin \delta \cos 2\alpha , \quad r = i\sin
\delta \sin 2\alpha ,\end{equation} $\delta = {\pi (n_o - n_e) h/
\lambda }$ is the optical thickness of the plate, $h$ is its
geometrical thickness, $\alpha $ is the orientation angle between
the optical axis of the plate and one of the basis, for example,
vertical direction.

Let us consider the action of the half-lambda plate on a
particular state $\Psi_\bot = \frac{1}{\sqrt 2
}{(|2,0\rangle-|0,2\rangle) } $, when the plate is oriented at
$22.5^\circ$. For the state $\Psi_\bot$ there are two nonzero
amplitudes $c_1 =c_3=1/\sqrt{2}$ and there is only one relative
phase $\varphi _{13} \equiv \varphi _1 - \varphi _3 = \pi $.
Taking into account that for a half-lambda plate $\delta = \pi
/2$, the corresponding transmission and reflection coefficients
are

\begin{equation}
t = r = \frac{i}{\sqrt 2 }.\end{equation}Thus the matrix $G$ has
the form

\begin{equation}
G = \left( {{\begin{array}{*{20}c}
 { - \frac{1}{2}} \hfill & { - \frac{1}{\sqrt 2 }} \hfill & { - \frac{1}{2}}
\hfill \\
 { - \frac{1}{\sqrt 2 }} \hfill & 0 \hfill & {\frac{1}{\sqrt 2 }} \hfill \\
 { - \frac{1}{2}} \hfill & {\frac{1}{\sqrt 2 }} \hfill & { - \frac{1}{2}}
\hfill \\
\end{array} }} \right).
\end{equation} Hence, acting by matrix $G$ on the state $\Psi _
\bot $ we get
\begin{equation}
G\Psi _ \bot = \frac{G}{\sqrt 2 }\left( {{\begin{array}{*{20}c}
 1 \hfill \\
 0 \hfill \\
 { - 1} \hfill \\
\end{array} }} \right) = \left( {{\begin{array}{*{20}c}
 0 \hfill \\
 { - 1} \hfill \\
 0 \hfill \\
\end{array} }} \right) = \Psi _2 .
\nonumber\end{equation} Note that such kind of transformations
cannot change the polarization degree of a qutrit. For the state
$\Psi_\bot$ chosen above, as well as for the state $\Psi_2$, the
polarization degree $P$ is zero.

In the experiment described below we used a simpler way to
generate qutrits. Biphotons were produced via collinear
frequency-degenerate spontaneous parametric down conversion in a
nonlinear crystal (BBO, type-I or type-II phase matching). For
type-I phase matching the polarization of both created photons was
vertical; i.e., the state $\Psi _3 $ was generated. Then, this
state was transformed using a quartz plate with a fixed optical
thickness. By changing the angle of the plate, the state $\Psi _3
= | {0_H ,2_V }\rangle $ is transformed according to the formula
$| {c_{in} } \rangle = G\Psi _3 $. For the case of type-II phase
matching the final state is $| {c_{in} } \rangle = G\Psi _2 $. Of
course the state $| {c_{in} } \rangle $ does not involve all
possible qutrit states because the transformation given by matrix
(7) preserves the polarization degree. Anyway using such a
transformation, we select some subset of qutrits to work with.

Such a simple method of the state preparation/transformation was
chosen in order to be able to compare the results of
reconstruction with the parameters of the input states, which
should be known with a high accuracy. The purpose of this work is
 the reconstruction of the initial state $|
{c_{in} } \rangle $.

We would like to emphasize that only pure qutrit states are
accessible by this method. To create a mixed state, some more
complicated method is to be used. This method allows one to create
arbitrary qutrit states and it implies a possibility to introduce
controlled delay between three fundamental states forming the
qutrit which could exceed the coherence length of the laser
pump~\cite{publish}.

\subsection{Coherence matrix}

We introduced only qualitative description of the qutrits based on
biphotons so far. The quantitative measure characterizing the
polarization properties of any single-mode state in the forth
moment in the field (including biphoton state) was proposed by
D.Klyshko in~\cite{kljetp}. It is a matrix consisting of six
fourth-order moments of the electromagnetic field. An ordered set
of such moments can be obtained using the direct product of
$2\times{2}$-coherence matrixes for both qubits. After normal
ordering, averaging, and crossing out the redundant row and column
the matrix takes the following form:

\begin{equation}K_4 \equiv \left( {{\begin{array}{*{20}c}
 A \hfill & D \hfill & E \hfill \\
 {D^ * } \hfill & C \hfill & F \hfill \\
 {E^ * } \hfill & {F^ * } \hfill & B \hfill \\
\end{array} }} \right).\end{equation}
The diagonal elements are formed by real moments, which
characterize the intensity correlation in two polarization modes
$H$ and $V$:

\begin{equation}A \equiv \langle {\hat{a}^{\dagger2}\hat{a}^2} \rangle ,
\quad B \equiv \langle {\hat{b}^{\dagger2}\hat{b}^2} \rangle ,
\quad C \equiv \langle {\hat{a}^\dagger \hat{b}^\dagger
\hat{a}\hat{b}} \rangle ,\end{equation} Nondiagonal moments are
complex:

\begin{equation} D \equiv \langle {\hat{a}^{\dagger2}\hat{a}\hat{b}}
\rangle , \quad E \equiv \langle {\hat{a}^{ \dagger2}\hat{b}^2}
\rangle , \quad F \equiv \langle {\hat{a}^\dagger\hat{b}^ \dagger
\hat{b}^2} \rangle. \end{equation} Three real moments (12) and
three complex ones (13) completely determine the state under
consideration. The elements of the matrix (11) are expressed
through the elements of the polarization density matrix. The
normalization condition,

\begin{equation}
A + B + 2C = 2, \end{equation} reduces the number of independent
real parameters, so for a mixed state we get 8 parameters as
expected. In the special case of a pure biphoton state, taking the
average in Eqs. (12,13) over the state (2), we obtain the matrix
components in the following form:

\begin{equation}
A = 2\left| {c_1 } \right|^2, \\
 B = 2\left| {c_3 } \right|^2, \\
 C = \left| {c_2 } \right|^2. \\
\end{equation}

\begin{equation} D =\sqrt 2 c_1^
* c_2 , \quad E = 2c_1^ * c_3, \quad F = \sqrt 2 c_2^
* c_3
\end{equation} So the links between the polarization density matrix
and the matrix (11) can be found comparing the corresponding
components of $(K_4)_{mk}$ and of $\rho \equiv | c \rangle \langle
c |$; $\rho _{mk} = c_m {c_k}^* ; m,k = 1,2,3$ for a pure state
and $\rho_{mk} = \overline {c_m {c_k}^*}$ for a mixed state where
the averaging, as usual, is taken over the classical probability
distribution. The statistics of the field is assumed to be
stationary and ergodic so the time-averaged values of the observed
quantities can be described in terms of a quantum statistical
ensemble. In this case $\langle {...} \rangle = Tr( {\rho ...} )$,
where $\rho $ is the polarization density operator.

\section{Methods of measurement}

What does it mean to measure the unknown state (2)? From the
experimental point of view, it means that the experimentalist has
to measure a complete set of real parameters (moments) determining
the state. To do this the state must be subject to a set of
unitary polarization transformations and projective measurements.
By doing this one picks out the outcomes, which are proportional
to the corresponding moments (12, 13) or their linear combination.
This procedure is known as quantum tomography. The quantum state
can be represented using either the wave function, density matrix,
or quasi-probability function (Wigner function). Probably the
correct way to use the term ``quantum tomography'' is only for the
reconstruction of the quasi-probability function because it gives
the graphical representation of the state as a 3D plot.
Nevertheless the term ``quantum tomography'' is also used for a
general procedure of complete state reconstruction. For a brief
review among the papers where this procedure was realized
experimentally, let us mention the works~\cite{smith,ariano,masal}
related to states defined by continuous variables. For states
characterized by discrete variables, such as two
polarization-spatial qubits, quantum tomography was realized
in~\cite{kwiat}. Recently quantum tomography has been performed
for orbital angular momentum entangled qutrits~\cite{lang} etc.

The physical idea behind the tomography procedure is performing
measurements of appropriately complete set of observables called
quorum~\cite{martini} or just ``looking'' at the state from
different positions. The minimal number of such positions might be
the number of real parameters determining the state.

According to Bohr's complementarity principle, it is impossible to
measure all moments (12,13) simultaneously, operating with a
single qutrit only. So to perform a complete set of measurements
one needs to generate a lot of representatives of a quantum
ensemble.

First of all, let us mention that at present, the only realistic
way to register single-mode biphoton field is using the
Brown-Twiss scheme. This scheme consists of a beam-splitter
followed by a pair of detectors connected with the coincidence
circuit. It means that registration of a single biphoton, which
carries the state (2), can give only a single event at the output
of the experimental set-up with some probability. So the
statistical treatment of the outcomes becomes extremely important.
For correlations between polarization degrees of freedom, which is
essential in the case under consideration, the Brown-Twiss scheme
must be accomplished with polarization filters introduced into
each arm.

\subsection{Qutrit tomography protocols}

We proposed two methods to perform polarization reconstruction of
a biphoton qutrit state $| c_{in}\rangle $.

\subsubsection{Protocol 1.}
The idea of the first method is splitting the state $| {c_{in} }
\rangle $ into two spatial modes and performing transformations
over two photons independently (Fig 3). These transformations can
be done using polarization filters placed in front of detectors.
Each filter consists of a sequence of quarter- and half-wave
plates and a polarization prism, which picks out definite linear
polarization, for example, the vertical one. A narrowband filter
centered at the doubled pump wavelength $\lambda = 2\lambda _p $
serves to make biphotons emitted from different sources
indistinguishable in frequency as well as to reduce the background
noise. An event is considered to be detected, if a pulse appears
at the output of the coincidence circuit. Approximately in half of
trials, one of the photons (signal, by convention) forming a
biphoton is going to one of the detectors, while the other one
(idler) is going to the other detector. In the remaining cases,
both photons appear in the same output beam-splitter arm, and
these events are not selected because they do not contribute to
coincidences.

\begin{figure}
\includegraphics[width=0.4\textwidth]{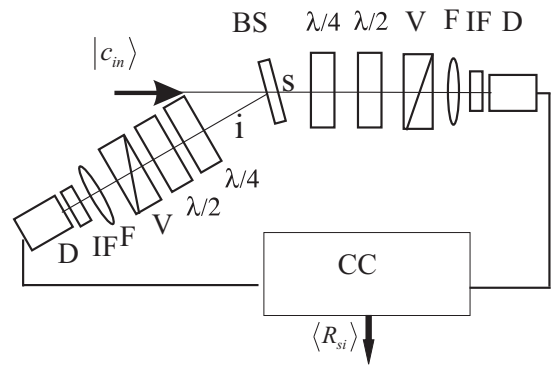}
\caption{Measurement block for Protocol 1. The Brown-Twiss scheme
for measuring intensity correlation between two polarization
modes. After spatial separation at the non-polarizing beam
splitter (\textbf{BS}), signal $(s)$ and idler $(i)$ photons
propagate through the quarter- and half wave plates, polarizing
prisms (\textbf{V}), focusing lenses (\textbf{F}) and interference
filters (\textbf{IF}) in two channels. Finally, photons are
registered by detectors (\textbf{D}). The coincidence rate from
the output of the coincidence circuit (\textbf{CC}) is
proportional to the fourth moment in the field $\langle {R_{si} }
\rangle $.}
\end{figure}

In the Heisenberg representation the polarization transformation for each
beam-splitter output port is given by:
\begin{eqnarray}
 \left( {{\begin{array}{*{20}c}
 {{a}'^\dagger} \hfill \\
 {{b}'^ \dagger} \hfill \\
\end{array} }} \right) = \left( {{\begin{array}{*{20}c}
 0 \hfill & 0 \hfill \\
 0 \hfill & 1 \hfill \\
\end{array} }} \right)
D_{\lambda /2} ( {\delta = \pi/2,\theta }) \nonumber\\
\times D_{\lambda/4} ( {\delta = \pi /4,\chi })
 \left( {{\begin{array}{*{20}c}
 {\frac{1}{\sqrt 2} } \hfill & 0 \hfill \\
 0 \hfill & {\frac{1}{\sqrt 2} } \hfill \\
\end{array} }} \right)
\left( {{\begin{array}{*{20}c}
 {a^\dagger} \hfill \\
 {b^ \dagger} \hfill \\
\end{array} }} \right).
\end{eqnarray} Four $2\times2$ matrixes in the right-hand side of Eq.
(17) describe the action of the non-polarizing beam-splitter,
$\lambda/4$-, $\lambda/2$- plates and vertical polarization prism
on the state vector of the signal (idler) photon;

\begin{equation}
D_{\lambda/2,\lambda/4} = \left( {{\begin{array}{*{20}c}
 t \hfill & r \hfill \\
 { - r^\ast } \hfill & {t^\ast } \hfill \\
\end{array} }} \right),
\nonumber\end{equation} where $r$ and $t$ - are the coefficients
introduced in Eq. (8), so for a $\lambda/4$-plate ($\delta =
\pi/4$),

\begin{subequations}
\begin{equation}
t_{\lambda/4} = \frac{1}{\sqrt 2}({1 + i\cos 2\chi } ),
r_{\lambda/4} = \frac{i}{\sqrt 2 }\sin 2\chi
\end{equation}
and for a
 $\lambda/2$-plate ($\delta = \pi/2$)

\begin{equation}
t_{\lambda/2} = i\cos ( {2\theta } ), r_{\lambda/2} = i\sin (
{2\theta } ). \end{equation} \end{subequations}Thus, there are
four real parameters (two for each channel) that determine
polarization transformations. Namely, these parameters are
orientation angles for two pairs of wave plates: $\theta _1 , \chi
_1 , \theta _2 , \chi _2 .$

As it was mentioned above, the output of the Brown-Twiss scheme is
the coincidence rate of the pulses coming from two detectors
$D_{s}$ and $D_{i}$. The corresponding moment of the fourth order
in the field has the following structure:

\begin{equation} R_{s,i} \propto \langle{ {{b'}_s}^\dagger{{b'}_i}^\dagger{b'}_s{b'}_i
}\rangle = R( {\theta _1 ,\chi _1 , \theta _2, \chi _2 }
)\end{equation} In the most general case this moment contains a
linear combination of six moments (12, 13) forming the matrix
$K_4$. So the main purpose of the quantum tomography procedure is
extracting these six moments from the set-up outcomes by varying
the four parameters of the polarization Brown-Twiss scheme.

Consider some special examples, which give the corresponding lines
in the complete protocol introduced below (Table I).
\begin{table*}
\caption{Protocol 1. Each line contains orientation of the half
($\theta_{s,i}$) and quarter ($\chi_{s,i}$) wave plates in the
measurement block. Last two columns show the corresponding moment
$R_n$ and the process amplitude $M_\nu(\nu =1,..9)$.}

\begin{tabular}{|c|c|c|c|c|c|c|c|} \hline
 & \multicolumn{4}{c|}{Parameters of the experimental set-up}
 & Moments to be measured & Amplitude of the
process\\\hline
  $\nu$&$\chi_s$&$\theta_s$&$\chi_i$&$\theta_s$&$R_{s,i}$&$M_\nu$\\\hline
  1. & 0 & $45^\circ$ & 0 & $-45^\circ$ & $A/4$ & $c_1/\sqrt{2}$\\\hline
  2. & 0 & $45^\circ$ & 0 & 0 & $C/4$ & $c_2/2$\\\hline
  3. & 0 & 0 & 0 & 0 & $B/4$ & $c_3/\sqrt{2}$\\\hline
  4. & $45^\circ$ & 0 & 0 & 0 & $1/8(B+C+2\textrm{Im}F)$ & $\frac{1}{2\sqrt{2}}c_2-\frac{i}{2}c_3$\\\hline
  5. & $45^\circ$ & $22.5^\circ$ & 0 & 0 & $1/8(B+C-2\textrm{Re}F)$ & $\frac{1}{2\sqrt{2}}c_2-\frac{1}{2}c_3$ \\\hline
  6. & $45^\circ$ & $22.5^\circ$ & 0 & $-45^\circ$ & $1/8(A+C-2\textrm{Re}D)$&  $\frac{1}{2}c_1-\frac{1}{2\sqrt{2}}c_2$\\\hline
  7. & $45^\circ$ & 0 & 0 & $-45^\circ$ & $1/8(A+C+2\textrm{Im}D)$& $\frac{1}{2}c_1-\frac{i}{2\sqrt{2}}c_2$ \\\hline
  8. & $-45^\circ$ & $11.25^\circ$ & $-45^\circ$ & $11.25^\circ$ & $1/16(A+B-2\textrm{Im}E)$ &  $\frac{1}{2\sqrt{2}}c_1+\frac{i}{2\sqrt{2}}c_3$\\\hline
  9. & $45^\circ$ & $22.5^\circ$ & $-45^\circ$ & $22.5^\circ$ & $1/16(A+B-2\textrm{Re}E)$ & $\frac{1}{2\sqrt{2}}c_1-\frac{1}{2\sqrt{2}}c_3$ \\ \hline
\end{tabular}

\end{table*}

First of all, it is obvious that for measuring real moments (12)
one needs to make polarization filters transmit both photon with
horizontal polarizations to measure $A$, both photons with
vertical polarization to measure $B$ and one photon with vertical
and another one with horizontal polarizations to measure $C$. To
do this all quarter-wave plates should be oriented at zero
degrees, then to install both half-wave plates at zero degrees for
measuring $B$; at $\theta _s = 45^\circ$ and $\theta _i =
45^\circ$ for measuring $A$; and at $\theta _s = 0^\circ$, $\theta
_i = 45^\circ$ for measuring $C$. These settings pick out the
squared modulus of corresponding amplitudes $c_3$, $c_1$ and
$c_2$.

The next example shows how to measure one of the complex moments
(13). To measure the real part of the moment $D$, let us set the
wave-plates in the Brown-Twiss scheme in the following way.

\noindent The idler channel:
\begin{subequations}
\begin{equation}
\lambda/4 :\chi _i=0^\circ, D_{\lambda/4} = \frac{1}{\sqrt 2
}\left( {{\begin{array}{*{20}c}
 {1 + i} \hfill & 0 \hfill \\
 0 \hfill & {1 - i} \hfill \\
\end{array} }} \right);
\end{equation}
\begin{equation}
\lambda /2: \theta_i= 45^\circ, D_{\lambda/2} = \left(
{{\begin{array}{*{20}c}
 0 \hfill & i \hfill \\
 i \hfill & 0 \hfill \\
\end{array} }} \right).
\end{equation}
\end{subequations} The signal channel:

\begin{subequations}
\begin{equation}
\lambda /4: \chi _s=45^\circ, D_{\lambda /4} = \frac{1}{\sqrt 2
}\left( {{\begin{array}{*{20}c}
 1 \hfill & i \hfill \\
 i \hfill & 1 \hfill \\
\end{array} }} \right);
\end{equation}
\begin{equation}
\lambda/2: \theta _s= 22.5^\circ, D_{\lambda/2} = \frac{1}{\sqrt 2
}\left( {{\begin{array}{*{20}c}
 i \hfill & i \hfill \\
 i \hfill & { - i} \hfill \\
\end{array} }} \right).
\end{equation}
\end{subequations}
 Substituting these matrices into Eq. (17) and taking
into account the commutation rules for the creation and
annihilation operators it is easy to get the final moment to be
measured:

\begin{equation}
R = \langle c |b_s^\dagger b_i^\dagger b_s b_i | c \rangle = 1/8(
{A + C - 2\textrm{Re}D} ). \nonumber\end{equation} A complete set
of the measurements called the tomography protocol is presented in
Table I. Each row corresponds to the setting of the plates to
measure the moment placed in the sixth column. The last one
corresponds to the amplitude of the process (see below).

This protocol was suggested and developed in~\cite{buropt,kriv}. A
similar protocol was considered in details earlier~\cite{kwiat}
 for estimating polarization state of a biphoton field,
generated in a frequency degenerate non-collinear mode. In this
case the biphoton field is represented as a pair of polarization
qubits.

Before describing the second method of the state measurement let
us make some remarks.

*We assume that the source generating qutrirs is stationary. Since
each measurement eliminates a qutrit one has to be sure that there
are a lot of copies of the initial state; each copy must be
prepared in the same quantum state. Such \textit{ensemble
approach} guarantees that the experimentalist deals with the same
quantum state in all trials. In other words, the outcomes provide
him with the information about the same quantum state and
elimination of a particular state does not affect the rest ones.

**The outcomes of the set-up are numbers related to the
corresponding moments (19). Usually this number is the coincidence
counting rate or the number of coincidences in a fixed time
interval. Due to the necessity of a proper normalization of the
state under investigation, the number of independent real
parameters grows up. The normalization is obtained from the
measurement of moments $A, B$ and $C$. Furthermore, only cosine
and sine of the phases $\varphi _{12} $ and $\varphi _{13}$ can be
measured in experiment as there is no way to measure the phases
directly. That is why the final number of real parameters to be
measured in experiment is 7 for a pure qutrit state and 9 for a
mixed state.

*** To minimize the errors caused by independent statistical fluctuations of
the outcomes, the number of moments (12, 13) entering in Eq. (19)
should be minimal.

\subsubsection{Protocol 2} In the second method of quantum tomography, a
biphoton-qutrit being measured is first subject to a sequence of
unitary transformations and, for each of such transformation, it
is fed to the Brown-Twiss scheme settled for measuring a fixed
moment. Using the wave plate with arbitrary optical thickness, one
can achieve the quorum varying the orientation of the plate $\mu
$.

In the most general case the coincidence counting rate in this
protocol is a periodic function of $\mu $, moreover, its Fourie
expansion contains nine harmonics of $\mu$ : $\cos(0\mu )$,
$\cos(2\mu )$, $\sin(2\mu )$, $\cos(4\mu )$, $\sin(4\mu )$,
$\cos(6\mu )$, $\sin(6\mu )$, $\cos(8\mu )$, $\sin(8\mu )$. These
harmonics depend linearly on the nine moments $A$, $B$, $C$,
$\textrm{Re}D$, $\textrm{Im}D$, $\textrm{Re}E$, $\textrm{Im}E$,
$\textrm{Re}F$, $\textrm{Im}F$. In other words, there is a
$9\times9$ matrix $T$ that links these nine harmonics to the nine
moments as shown below.

\begin{equation}
\left( {{\begin{array}{*{20}c}
 {\cos 0\mu } \hfill \\
 {\cos 2\mu } \hfill \\
 {\sin 2\mu } \hfill \\
 {\cos 4\mu } \hfill \\
 {\sin 4\mu } \hfill \\
 {\cos 6\mu } \hfill \\
 {\sin 6\mu } \hfill \\
 {\cos 8\mu } \hfill \\
 {\sin 8\mu } \hfill \\
\end{array} }} \right) = T\left( {{\begin{array}{*{20}c}
 A \hfill \\
 B \hfill \\
 C \hfill \\
 {\textrm{Re}F} \hfill \\
 {\textrm{Im}F} \hfill \\
 {\textrm{Re}D} \hfill \\
 {\textrm{Im}D} \hfill \\
 {\textrm{Re}E} \hfill \\
 {\textrm{Im}E} \hfill \\
\end{array} }} \right).
\nonumber\end{equation}Unfortunately, the inverse matrix does not
always exist. To simplify the problem we put only a single wave
plate with fixed optical thickness ($\delta _s = \pi/4,\delta _i =
\pi/2$) and fixed orientation $\chi_s ,\theta_i $ in each channel
of the Brown-Twiss scheme. In order to make sure the inverse
matrix exists one needs to maximize the determinant of the matrix
$T$ over the orientations of the plates $\chi _s ,\theta _i $.
After accomplishing this procedure we obtain $\chi _s \approx
19^\circ$, $\theta _i \approx - 28.5^\circ$ (Fig.4).

\begin{figure}
\includegraphics[width=0.4\textwidth]{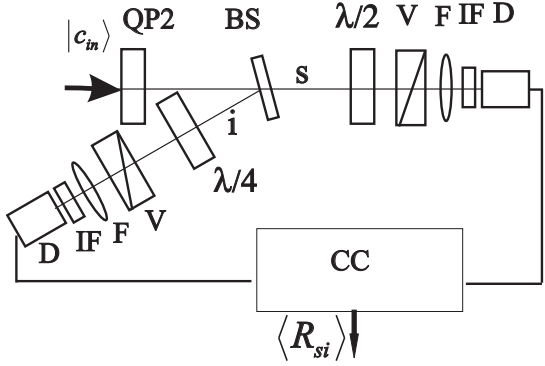}
\caption{Measurement block for Protocol 2. Additional control
quartz plate (\textbf{QP2}) serves as state $| {c_{in} } \rangle $
tomography transformer. Only a single wave plate is introduced in
each channel.}
\end{figure}

Instead of finding the links between the harmonics and moments,
there is a more elegant method to reconstruct the quantum state
using the second protocol (see section III.D). This method is
considered in the present work for the first time.

\subsection{Experimental implementation. Protocol 1}

The experimental set-up for the quantum tomography of qutrits
using protocol 1 is shown in Fig.5. The \textit{preparation block}
includes a 2-mm BBO crystal with either type-I or type-II
degenerate and collinear phase-matching, which is pumped with
cw-argon laser operated at 351nm wavelength. In the case of
type-II phase-matching, an additional quartz compensator is
introduced right after the crystal. The state $\Psi _3 = | {0,2}
\rangle$ (for type-I) or $\Psi _2 = | {1,1} \rangle $ (for
type-II) generated in the crystal is fed to the
\textit{transformation block}. This block consists of the quartz
plate with fixed optical thickness $\delta = \mbox{0.9046}$ and
variable orientation $\alpha$. So the state, which is to be
measured, is determined by the parameter $\alpha $. The
\textit{measurement block} is a Brown-Twiss scheme equipped with
polarization filters placed in both arms (Fig.3). Pulses coming
from a couple of single-photon modules (EG{\&}G SPCM-AQR) were fed
to the counter through a standard time-to-amplitude converter.

\begin{figure}[b]
\includegraphics[width=0.45\textwidth]{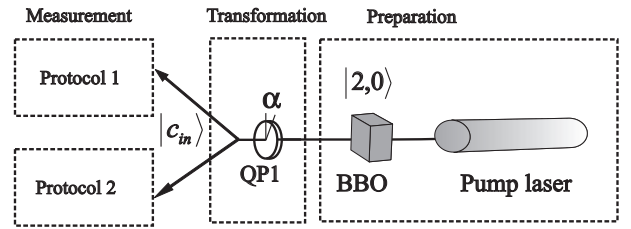}
\caption{Scheme of qutrits tomograph, consisting of three blocks.
Preparation block includes pump laser(s) and nonlinear crystal(s).
Transformation block is the quartz plate (\textbf{QP1}) which
orientation angle $\alpha $ determines the final state to be
measured. The measurement block depends on the protocol to be used
(see Fig.3 and Fig.4).}
\end{figure}

In our experiments the exposure time for measuring each moment is
5 sec. This time is an important experimental parameter. Each
measurement consists of 30 runs, after which the scheme is reset.
Namely each measurement is performed by setting the angles of wave
plates $\chi_j$ and $\theta_j $ in both arms according to the
tomographic protocol (Table I). After 30 runs, a new set of angles
is selected and the next moment is measured in the same way. The
output data of the set-up are the mean coincidence counting rates.
Examples of behavior for some moments ($A, B, C, \textrm{Re}F,
\textrm{Im}F$) versus the orientation of the plate \textbf{QP1}
are plotted in Fig.6.

\begin{figure}
\includegraphics[width=0.4\textwidth]{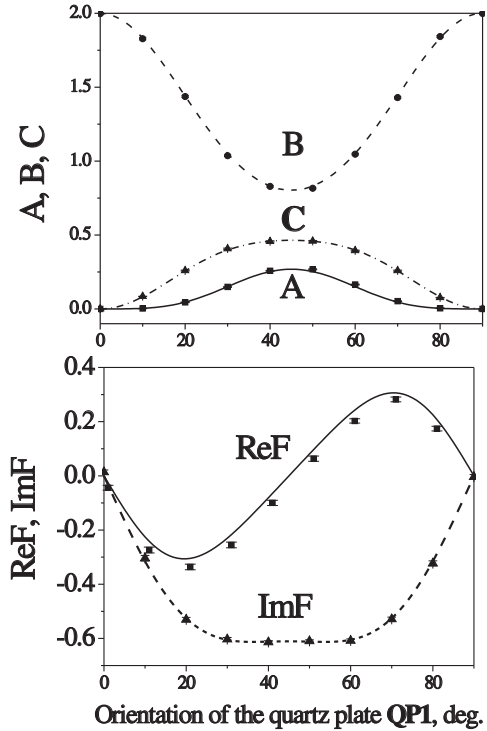}
\caption{Some components of the matrix $K_4$ versus the
orientation of the quartz plate \textbf{QP1}. Different angles of
the plate correspond to different states sent to the measurement
block. The plot at the top corresponds to measured real moments
$A$ (squares), $B$ (circles), $C$ (triangles) and theoretical
predictions $A (-), B(--), C (-\circ-)$. The plot at the bottom
shows measured complex moment $\textrm{Re}F$ (squares),
$\textrm{Im}F$ (triangles) and theoretical predictions
$\textrm{Re}F$ (-), $\textrm{Im}F$ (--).}
\end{figure}

\subsection{Experimental implementation. Protocol 2}

For the second method we used Ti:Sa laser with pulse duration
about 250 fsec, operating at 800nm. After frequency doubling, the
UV radiation with 400nm wavelength was sent into the same set-up
as described above. For this protocol we used 2-mm BBO crystal cut
for collinear degenerate type-I phase-matching. A quartz plate
with the optical thickness $\delta = 0.656$ is placed after the
BBO crystal to prepare the qutrit state to be measured. The
measurement part of the set-up was slightly changed (Fig.4).
Additional control quartz plate introduced in front of the beam
splitter accomplishes the protocol. Its orientation angle $\mu $
is a parameter defining the measurement process. The control plate
is rotated with 5 degrees step from 0 up to 360 degrees so that
Protocol 2 consists of 72 measurements. Each arm of the
Brown-Twiss scheme contains either quarter or half-wave plate with
fixed orientation. The orientations are $\chi _1 = 18.8^\circ$ for
the quarter-wave plate in the first channel and $\theta _2 =-
28.5^\circ$ for the half-wave plate in the second channel.
Protocol 2 is easier to implement since only a single parameter
$\mu $ is changed whereas four $\chi _{1,2} ,\theta _{1,2} $
parameters are varied in the Protocol 1. In perspective, this kind
of protocol allows one to automate the quantum tomography
procedure: the control plate can be rotated continuously and the
reconstruction of the quantum state can be based on the analysis
of coincidence rates corresponding to the respective values of
$\mu _{i} (i= 1,..,72)$.

\subsection{Statistical reconstruction of biphoton-field qutrits
from the outcomes of mutually complementary measurements}

Each of the 9 processes from Protocol 1 as well as of the 72
processes from Protocol 2 is described by its amplitude $M_\nu $.
From the statistical point of view, the squared modulus of the
process amplitude specifies the intensity of the event generation:

\begin{equation}
R_\nu = M_\nu ^\ast M_\nu \end{equation} The considered processes
are examples of mutually complementary sets of measurements in the
sense of Bohr's complementarity principle. The event-generation
intensities $R_\nu $ for both protocols are the main quantities
accessible from the measurement. Making the bridge between
statistical and physical description of the process the quantities
$R_\nu $ coincide with the fourth moments in the field introduced
above in Eq. (19). Their dimension is frequency unit (Hz). The
number of events occurring within any given time interval obeys
the Poisson distribution. Therefore, the quantities $R_\nu $
specify the intensities of the corresponding mutually
complementary Poisson processes and serve as estimates of the
Poisson parameters $\lambda _\nu $ (see below).

Although the amplitudes of the processes cannot be measured
directly, they are of the greatest interest as these quantities
describe fundamental relationships in quantum physics. It follows
from the superposition principle that the amplitudes are linearly
related to the state-vector components. So the main purpose of
quantum tomography is reproduction of the amplitudes and state
vectors, which are hidden from direct observation.

The linear transformation of the state vector $c={ \{c_1, c_2,
c_3\}}$ into the amplitude of the process $M$ is described by a
certain matrix $X$. For example, considering the first protocol
this matrix can be easily obtained from Table I (last column in
Table I):

\begin{equation} X = \left( {{\begin{array}{*{20}c}
 {1/\sqrt 2 } \hfill & 0 \hfill & 0 \hfill \\
 0 \hfill & {1/2} \hfill & 0 \hfill \\
 0 \hfill & 0 \hfill & {1/\sqrt 2 } \hfill \\
 0 \hfill & {1/(2\sqrt2 )} \hfill & {
- i/2}\hfill \\
 0 \hfill & {1/(2\sqrt 2)} \hfill & {
- 1/2} \hfill \\
 {1 /2}
\hfill & { - 1 /(2\sqrt 2) } \hfill & 0
\hfill \\
 {1 /2}
\hfill & { - i/ (2\sqrt 2)} \hfill & 0
\hfill \\
 {1/(2\sqrt 2) }  \hfill & 0 \hfill &
{i/(2\sqrt 2 )} \hfill \\
 {1 /(2\sqrt 2 )} \hfill & 0 \hfill &
{ - 1/ (2\sqrt 2)} \hfill \\
\end{array} }} \right)
\end{equation}Using the matrix $X$ the complete set of nine
amplitudes of the processes can be expressed by a single equation,
\begin{equation}
Xc = M \end{equation} The matrix $X$ is an instrumental matrix for
a set of mutually complementary measurements, by analogy with the
conventional instrumental function. The implementation of the
method to the first protocol has been considered
in~\cite{boglett}.

Consider an algorithm allowing one to calculate the instrumental
matrix $X$ for Protocol 2. The matrix consists of 72 rows (the
number of control plate orientations) and 3 columns (the dimension
of Hilbert space for qutrits). Each row is formed in the following
way. Using coefficients $t^{(s,i)}$ and $r^{(s,i)}$ of the wave
plates introduced to the signal and idler channels of the
Brown-Twiss scheme (18a,b) the three-element row, which defines
the process amplitude right after the control plate, can be
written in the form

\begin{equation}
l = \left[
r^{(s)}_{\lambda/4}r^{(i)}_{\lambda/2}\quad\frac{1}{\sqrt{2}}(r^{(s)}_{\lambda/4}t^{(i)}_
{\lambda/2}+r^{(i)}_{\lambda/2}t^{(s)}_{\lambda/4})\quad
t^{(s)}_{\lambda/4}t^{(i)}_{\lambda/2}\right].
\end{equation} The unitary matrix $G$ is defined by the control
plate according to (7, 8), with a replacement $\alpha \to \mu $,
where $\mu $ is the control plate orientation (it takes $72$
values from $0^\circ$ to $355^\circ$ ). We chose the control plate
to be a quarter-wave plate, so $\delta = \pi /4$. Finally,
\begin{equation}
G = G\left( {\mu _i } \right),i = 1,2,...,72. \end{equation} Each
row of the instrumental matrix $X$ (72 rows, 3 columns) is defined
by the product of the row $l$ (which is the same for any process)
and the matrix $G$ (which is defined by the control plate
orientation angle)

\begin{equation}
X_i = lG\left( {\mu _i } \right),i = 1,2,...,72, \end{equation}
where $X_{i}$ is the $i$th row of the matrix $X$.

\section{Methods of quantum state reconstruction }

In the simplest case the density matrix can be estimated directly
from the measurements. Since the set of experimental data is
limited in this case, the reconstructed density matrix may have
non-physical properties like negative eigenvalues. But in the
general case of $s-$dimensional systems the problem of density
matrix reconstruction using direct results of measurements can not
be solved since the corresponding inverse problem is ill-posed.

When analyzing the experimental data, we use the so-called root
estimator of quantum states~\cite{bogarxiv}. This approach is
designed specially for the analysis of mutually complementary
measurements (in the sense of Bohr's complementarity principle).
The advantage of this approach consists of the possibility of
reconstructing states in a high- dimensional Hilbert space and
reaching the accuracy of reconstruction of an unknown quantum
state close to its fundamental limit. Below we consider two
methods of quantum state root estimation that give similar
results. They are the least-squares method (LSM) and
maximum-likelihood method (MLM).

\subsection{Least-squares method}

In statistical terms, Eq. (24) is a linear regression equation. A
distinctive feature of the problem is that only the absolute value
of the process amplitude $M$ is measured in the experiment. The
estimate of the absolute value of the amplitude is given by the
square root of the corresponding experimentally measured
coincidence rate:

\begin{equation}
| {M_\nu } |^{\exp } = \sqrt {k_\nu /t} ,\end{equation}

\noindent
where $k_\nu $ is the number of events (coincidences) detected in the $\nu
$-th process during the measurement time $t$.

It is worth noting that, by the action of the root-square
procedure on a Poissonian random value, one gets at the random
variable with a uniform variance, i.e., at the variance
stabilization~\cite{cram}. Note also, since we deal not with event
probabilities but with their rates or intensities, it is
convenient to use un-normalized state vectors. These vectors allow
the coincidence counting rate (event-generation intensities) to be
derived directly from Eqs. (22), (24) without introducing the
coefficients related to the biphoton generation rate, detector
efficiencies, etc. The dimensionality of the vector state obtained
in such a way is $1 / \sqrt {\textrm{time}} $. The final state
vector obtained by the reconstruction procedure, nevertheless,
should be normalized to unity.

Assuming that the variances of different $| {M_\nu } |^{\exp }$
are independent and identical, one can apply the standard least-
squares estimate to Eq. (24)~\cite{cram}:

\begin{equation}
c = ( {X^\dagger X} )^{ - 1}X^\dagger M \end{equation} Unlike the
traditional least-squares method, Eq. (29) cannot be used for the
explicit estimation of the state vector $c$, because it is to be
solved by the iteration method. The absolute value of $M$ is known
from the experiment $ ( |M_\nu| = | M_\nu |^{\exp } )$. We assume
that the phase of vector $Xc$ at the $i$-th iteration step
determines the phase of the vector $M$ at the $i+1$-th step. In
other words the phase is determined by the iteration procedure.

It turns out that, for the Gaussian approximation of Poisson's
quantities, this least-squares estimate coincides with a more
exact and rigorous maximum-likelihood estimate considered below.

\subsection{Maximum-likelihood method}

The likelihood function is defined by the product of Poissonian
probabilities:

\begin{equation}
L = \prod\limits_i {\frac{( {\lambda _i t_i })^{k_i }}{k_i !}} e^{
- \lambda _i t_i }, \end{equation} where $k_{i}$ is the number of
coincidences observed in the $i$-th process during the measurement
time $t_{i}$, and $\lambda _i $ are the unknown theoretical
event-generation intensities (expected number of coincidences
proportional to the moments in the field), whose estimation is the
subject of this section.

The logarithm of the likelihood function is, if we omit an insignificant
constant,
\begin{equation}
\ln L = \sum\limits_i {( {k_i \ln ( {\lambda _i t_i } ) - \lambda
_i t_i } )} . \end{equation} Let us introduce the matrices with
the elements defined by the following formulas:

\begin{equation}
I_{js} = \sum\limits_i {t_i X_{ij}^\ast X_{is} } , \end{equation}

\begin{equation}J_{js} = \sum\limits_i {\frac{k_i }{\lambda_i}X_{ij}^\ast X_{is}
};  \quad j,s = 1,2,3. \end{equation} The matrix $I$ is determined
from the experimental protocol and, thus, is known \textit{a
priori }(before the experiment). This is the Hermitian matrix of
Fisher's information. The matrix $J$ is determined by the
experimental values of $k_{i}$ and by the unknown event-generation
intensities $\lambda _i$. This is the empirical matrix of Fisher's
information (see also Appendix).

In terms of these matrices, the condition for the extremum of the
function (31) can be written as

\begin{equation}
Ic = Jc.\end{equation}  Hence, it follows that

\begin{equation}
I^{ - 1}Jc = c. \end{equation} The latest relationship is known as
the likelihood equation. This is a nonlinear equation, because
$\lambda _i$ depends on the unknown state vector $c$. Because of
the simple quasi-linear structure, this equation can easily be
,solved by the iteration method~\cite{bogarxiv}. The quasi-
identity operator $I^{ - 1}J$ acts as the identical operator upon
only a single vector in the Hilbert space, namely, on the vector
corresponding to the solution of Eq. (35) and representing the
maximum possible likelihood estimate for the state vector. The
condition for the existence of the matrix $I^{ - 1}$ is a
condition imposed on the initial experimental protocol. The
resulting set of equations automatically includes the
normalization condition, which is written as
\begin{equation}
\sum\limits_i {k_i } = \sum\limits_i {( {\lambda _i t_i } )} \quad
. \end{equation} This condition implies that, for all processes,
the total number of detected events is equal to the sum of the
products of event detection frequencies during the measurement
time.

\subsection{ Analysis of the experimental data}

\subsubsection{Pure state reconstruction}

The examples of qutrit state reconstruction using both the
least-squares and maximum-likelihood methods are given in Table
II.

\begin{table*}[t]
\caption{Results of the state reconstruction. The left column
indicates the orientation of the quartz plate \textbf{QP1},
determining the state to be measured. Values of the optical
thickness of \textbf{QP1}  are $\delta= 0.656$ for the pulsed
regime (Protocols 1, 2), and $\delta = 0.9046$ for the cw regime
(Protocol 1). Theoretical state vectors are placed in the right
column. The table contains the amplitudes of the reconstructed
states $(c_1, c_2, c_3)$ as well as their fidelities, calculated
by least-squared (LSM) and maximum-likelihood (MLM) methods.}
\begin{tabular}{|c|c|c|c|c|c|} \hline
 \multicolumn{6}{|c|}{Pulsed regime, $\delta=0.656$, Protocol 1}\\\hline
 $\alpha$& \multicolumn{2}{c|}{Fidelity}
 & \multicolumn{2}{c|}{State vector: experiment} & State vector: theory\\
 & LSM & MLM & LSM & MLM & $(c_1,c_2,c_3)_{theory}$ \\\hline
$0^\circ$&0.99981&0.99979&-0.0046+0.0040i&-0.0065+0.0057i&0\\
 & & &-0.0050-0.0115i&-0.0053-0.0102i&0\\
 & & &0.9999&0.9999&1\\\hline
$40^\circ$&0.9989&0.9989&-0.3669-0.0691i&-0.3669-0.0687i&-0.3482-0.0948i\\
 & & &-0.0657+0.6814i&-0.0653+0.6815i&-0.0900+0.6732i\\
 & & &0.6261&0.6261&0.6392\\\hline
$80^\circ$&0.9993&0.9993&-0.0088+0.0439i&-0.0091+0.0439i&-0.0136+0.0413i\\
 & & &0.1691+0.2587i&0.1697+0.259i&0.1691+0.2338i\\
 & & &0.9500&0.9498&0.9565\\\hline
\multicolumn{6}{|c|}{Pulsed regime, $\delta=0.656$, Protocol
2}\\\hline
$0^\circ$&0.99846&0.99847&-0.0071-0.0135i&-0.0072-0.0135i&0\\
 & & &0.0359+0.0046i&0.0357+0.0046i&0\\
 & & &0.9992&0.9992&1\\\hline
$40^\circ$&0.9991&0.9991&-0.3442-0.1139i&-0.3444-0.1142i&-0.3482-0.0948i\\
 & & &-0.0987+0.6546i&-0.0990+0.6545i&-0.0900+0.6732i\\
 & & &0.6560&0.6559&0.6392\\\hline
$80^\circ$&0.9981&0.9981&-0.0093+0.0430i&-0.0094+0.0430i&-0.0136+0.0413i\\
 & & &0.2122+0.2408i&0.2121+0.2408i&0.1691+0.2338i\\
 & & &0.9461&0.9461&0.9565\\\hline
\multicolumn{6}{|c|}{CW-regime, $\delta=0.9046$, Protocol
1}\\\hline
$0^\circ$&0.99325&0.99313&-0.0030-0.0512i&-0.0028-0.0514i&0\\
 & & &0.9966&0.9966&1\\
 & & &-0.0015-0.0642i&-0.0013-0.0649i&0\\\hline
$60^\circ$&0.9886&0.9799&0.7236&0.7244&0.7052\\
 & & &0.1165-0.1231i&0.1245-0.1210i&0.0392-0.0616i\\
 & & &0.2792+0.6080i&0.1694+0.6453i&0.2990+0.6387i\\\hline

\end{tabular}
\end{table*}

The value of the fidelity parameter $F$ is defined as

\begin{equation}
F = | \langle c_{theory}| c_{exp}\rangle|^2 . \end{equation} It
gives the conventional measure of correspondence between the
theoretical and experimental state vectors.

The dependence of fidelity on the amount of experimental data
obtained is shown in Fig.7. This figure shows the fidelity
achieved in the experiment in comparison with the theoretical
range (see Appendix for more details). The lower boundary
corresponds to 5{\%} quantile of statistical distribution, while
the upper to 95{\%}-- quantile. It is clearly seen that the
fidelity value achieved experimentally for a small volume of
experimental data is completely within the limits of the
theoretical range, while it goes out for a higher volume. Such
behavior of fidelity is due to the existence of two different
error types arising under the reconstruction of quantum states.
Let us call them statistical and instrumental errors,
respectively. The statistical errors are caused by a finite number
of quantum system representatives to be measured. As the
measurement time increases, the information about the quantum
state of interest progressively increases (see Appendix).
Accordingly, the statistical error becomes smaller. The
instrumental errors are caused by the researcher's incomplete
knowledge of the system; i.e., more exact information exists, in
principle, but it is inaccessible to the experimenter. Thus, a
comparison between the state reconstruction result and the
fundamental statistical level of accuracy can serve as a guide for
the parameter adjustment of the set-up.
\begin{figure}
\includegraphics[width=0.5\textwidth]{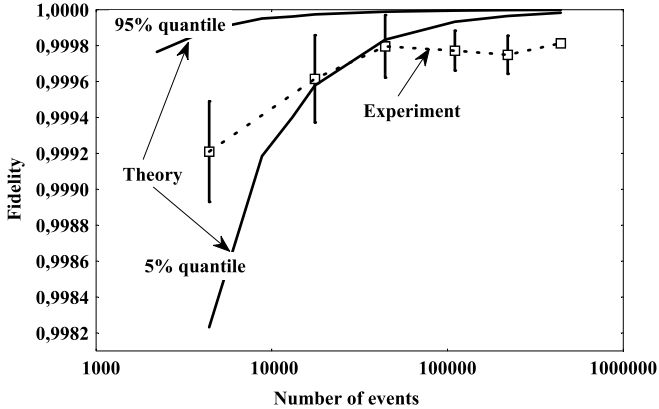}
\caption{Fidelity dependence on the sample size. Mean values and
standard deviations corresponding to the sample volumes $f$ =
0.01; 0.04; 0.1; 0.25; 0.5; 1.0.}
\end{figure}

Thus, for a small volume of experimental data, statistical errors
prevail, whereas for large sample sizes, the setting errors and
the instability of protocol parameters dominate. The number of
events at which the statistical error becomes smaller than the
instrumental error can be called the coherence volume. Numerically
coherence volume can be estimated as the intersection point
between the experimental fidelity and the lower theoretical
fidelity curve. In our case this value is about 25,000-30,000
events. Starting approximately from this value, fidelity is
reaching saturation and further growth of experimental data volume
does not lead to an increase in the precision of quantum system
estimation.

Fig.7 relates to the state defined by orientation angle of quartz
plate \textbf{QP1} $\alpha = 50^\circ$ (for Protocol 2). To plot
Fig.7 we used the following technique for passing from full-volume
experiment to a partial-volume experiment. Let us consider the
parameter $0 < f \le 1$ that characterizes the volume of
experimental data. Suppose that $f = 1$ for a full-volume
experiment. A partial volume experiment may be introduced
considering the observation time ${t}'_\nu = ft_\nu $ instead of
$t_\nu $. Hence, performing a single full-volume experiment means
providing with a large (practically infinite) number of
partial-volume experiments.

For a given volume of experimental data $f$ each event from the
full-volume experiment is picked up with the probability $f$ and
rejected with the probability $1 - f$. Due to the presence of
statistical fluctuations the equation for the number of
observations, $k_\nu ( t'_\nu ) = fk_\nu ( t_\nu )$, is violated.
Therefore a unique estimate of the state vector corresponds to
every partial-volume experiment. Fig.7 shows mean values and
standard deviations corresponding to volumes $f =$ 0.01; 0.04;
0.1; 0.25; 0.5; 1.0. For each $f < 1$ ten experiments were
simulated.

The results of informational fidelity research, introduced in
Appendix, are shown in Figs.8, 9. These figures correspond to the
same data set as shown in Fig.7. Distribution density of
informational fidelity for a small (compared to the coherence
volume) sample size closely agrees with the theoretical result
given by (A14) (see Fig.8). In this case the instrumental error is
negligibly small compared to the statistical one. When the sample
volume is close to the coherence volume (Fig. 9) the influence of
instrumental and statistical errors is about equal. In other
words, the informational losses caused by averaging over
instrumental errors are approximately equal to the losses caused
by statistical ones. Finally, if the sample size is greater than
the coherence volume, instrumental errors predominate. It means
that the statistical informational errors are negligibly small
compared to the instrumental ones.

\begin{figure}
\includegraphics[width=0.5\textwidth]{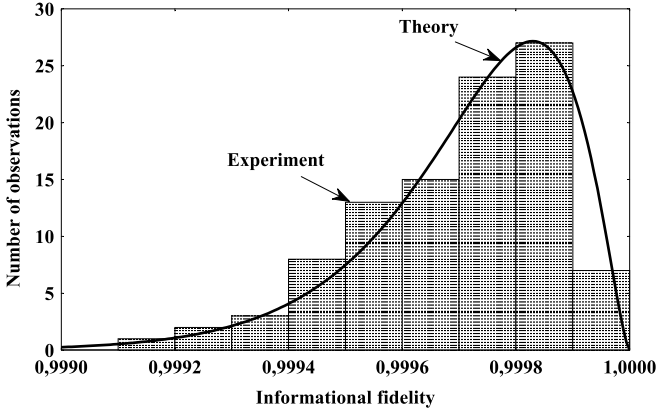}
\caption{ Informational  $\chi^2$   criterion for small sample
sizes:sample size=4400.}
\end{figure}
\begin{figure}
\includegraphics[width=0.5\textwidth]{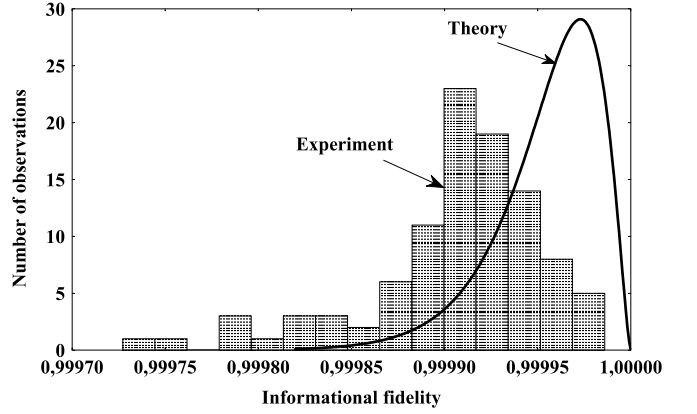}
\caption{Informational  $\chi^2$   criterion for large sample
sizes: sample size=27750. The disagreement between observations
and theoretical curve for large sample sizes is due to the
instrumental error.}
\end{figure}

\subsubsection{Mixture separation algorithm}

Let us describe the algorithm for reconstructing a two-component mixed
state. This algorithm can be easily generalized to an arbitrary number of
components.

The total number of events observed in every process is divided between the
components proportional to the intensity,

\begin{equation}
k_\nu ^{(1)} = k_\nu \frac{\lambda _\nu ^{(1)} }{\lambda _\nu
^{(1)} + \lambda _\nu ^{(2)} }, k_\nu ^{(2)} = k_\nu \frac{\lambda
_\nu ^{(2)} }{\lambda _\nu ^{(1)} + \lambda _\nu ^{(2)}
}\end{equation} where  $\nu = 1,2,...,\nu _{max}$ and $\nu _{\max
} $ is the total number of processes, $\lambda _\nu ^{(1)} $ and
$\lambda _\nu ^{(2)} $ are the estimates of intensities of
processes for a given step of the iteration procedure.

At a certain iteration step, let us represent $k_\nu $ as a sum of two
components,

\begin{equation} k_\nu = k_\nu ^{(1)} + k_\nu ^{(2)} .
\end{equation} For each component, we can obtain the estimates for
the state vector, amplitudes, and intensity of the processes
according to the method of pure state analysis described in the
previous section. Since we get new intensity estimates, let us
again split the total number of events in every process
proportionally to the intensities of the components. In such a
way, a new iteration is formed and the whole procedure is
repeated. The described process is called quasi-Bayesian
algorithm~\cite{bogarxiv}.

As a result, the iteration process converges to some
(non-normalized) components $c^{(1)}$ and $c^{(2)}$. Thus, the
mixture separation algorithm reduces to numerous estimations of
pure components according to the simple algorithm described above
in section IV.B. As a result of the whole algorithm execution, the
estimate for the density matrix of the mixture appears:

\begin{equation}\rho = c^{(1)}c^{(1)\dagger} + c^{(2)}c^{(2)\dagger},
\end{equation}

\begin{equation}
\rho \to \frac{\rho }{\textrm{Tr}\left( \rho \right)}.
\end{equation}The last procedure is normalization of
the density matrix.

A remarkable feature of the algorithm is that according to
numerical calculations, independent on zero-approximation
selection of the mixture components, the resulting density matrix
$\rho $ is always the same. Of course, the components $c^{(1)}$
and $c^{(2)}$ are different for the random selection of the
zero-approximation.

The mixed state reconstruction accuracy is described by the following
fidelity:
\begin{equation}
F = \left[ \textrm{Tr}\sqrt {\sqrt {\rho ^{( 0 )}} \rho \sqrt
{\rho ^{(0)}} }  \right]^2, \end{equation} where $\rho ^{(0)}$ and
$\rho $ are the exact and reconstructed density matrices
respectively. For a pure state $( {\rho ^2 = \rho ,( \rho
^{(0)})^2 = \rho ^{(0)}})$ fidelity (42) converts to (37).

Actually in the present work we did not intend to generate a given
mixed state of qutrit in experiment, it will be done later
~\cite{publish}. Nevertheless, applying the described algorithm to
the data we can check whether the state produced in our system is
pure. For example, consider the case when the state $\Psi _2 = |
{1,1} \rangle $ is fed to the quartz plate \textbf{QP1} (see Table
II). This state is the most interesting to be tested, since $|
{H,V} \rangle $ and $| {V,H} \rangle $ are distinguishable due to
the polarization dispersion in BBO crystal. Namely,
extraordinarily polarized photons $(H)$ propagate faster than
ordinary $(V)$ ones in the crystal. Therefore a group velocity
compensator has to be used for making them
indistinguishable~\cite{shih}. Non-perfect compensation (we
reached 95{\%} visibility for polarization interference) is the
main reason why the fidelity reconstruction for these states is
not so high. The results of applying quasi-Bayesian algorithm to
the reconstructed state are in Table III. We chose the state
corresponding to the angle $\alpha = 30^\circ$. It is clearly seen
that the weight of the first principal component is much greater
that of the second one. Doing the same procedure with the $\Psi _1
=| {2,0} \rangle $ initial state, we have checked that the
estimator for a pure state vector is extremely close to the
estimator of the major density matrix component.

\begin{table}
\caption{Example of the mixture separation using quasi-Bayesian
algorithm for the given state. CW-regime, protocol 1.}

\begin{tabular}{|c|c|c|} \hline
State vector:theory & \multicolumn{2}{c|}{Density matrix:
experiment}\\\hline
&First principal&Second principal \\
$\alpha=30^\circ$& component &component \\
$\delta=0.9046$&weight=0.9238&weight=0.0762\\
$(c_1,c_2,c_3)_{theory}$&$(c_1,c_2,c_3)^1_{exp}$&$(c_1,c_2,c_3)^2_{exp}$\\\hline
0.7052&0.7019&-0.3027-0.2858i\\
-0.0392-0.0616i&-0.0466-0.1325i&-0.6529+0.3291i\\
0.2990-0.6387i&0.2245-0.6612i&0.5140-0.1668i\\\hline

\multicolumn{3}{|c|}{Fidelity=0.9916}\\\hline
\end{tabular}
\end{table}
To illustrate the quasi-Bayesian approach, let us consider a
result of reconstruction for a two-component mixture using
protocol 1. Suppose one has a mixture of two pure states prepared
from $| {2,0} \rangle $ by quartz plates \textbf{QP1'} and
\textbf{QP1''} oriented at angles $\alpha = - 30^0$ and $\alpha =
50^0$, respectively. Let the optical thickness of both plates be
$\delta = 0.656$. Ten thousands events were generated (on the
average) for every component. The theoretical density matrix for
the mixed state under consideration is
\begin{widetext}
\begin{equation}\rho ^{(0)} = \left(
{\begin{array}{*{20}c}
 {0.1134} \hfill & {0.0263 + 0.0808i} \hfill & { - 0.1987 - 0.0558i} \hfill
\\
 {0.0263 - 0.0808i} \hfill & {0.4404} \hfill & {0.0679 + 0.0752i} \hfill \\
 { - 0.1987 + 0.0558i} \hfill & {0.0679 - 0.0752i} \hfill & {0.4462} \hfill
\\
\end{array} }\right)
\end{equation}

A typical example of a reconstructed density matrix is the
following:

\begin{equation}
\rho = \left( {\begin{array}{*{20}c}
 {0.1162} \hfill & {0.0294 + 0.0808i} \hfill & { - 0.1965 - 0.0691i} \hfill
\\
 {0.0204 - 0.0808i} \hfill & {0.4298} \hfill & {0.0697 + 0.0796i} \hfill \\
 { - 0.1965 + 0.0691i} \hfill & {0.0697 - 0.0796i} \hfill & {0.4540} \hfill
\\
\end{array} } \right)
\end{equation}
\end{widetext}
The reconstructed matrix fidelity is $F = 0.999431$. Analysis of
the principal components of density matrix is given in Table IV.

\begin{table}
\caption{Analysis of the principal components of the density
matrix for the state (43, 44): numerical simulation.}
\begin{tabular}{|c|c|c|} \hline
\multicolumn{2}{|c|}{State vector}&Fidelity\\
\multicolumn{2}{|c|}{$(c_1,c_2,c_3)$}&\\\hline
\multicolumn{3}{|c|}{First
principal component}\\\hline

Experiment&Theory&\\
weight=0.6188&weight=0.6143&\\\hline
-0.3658-0.0448i&-0.3668-0.0211i&0.9985\\
0.2085+0.4743i&0.2294+0.4934i&\\
0.7718&0.7543&\\\hline \multicolumn{3}{|c|}{Second principal
component}\\\hline
Experiment&Theory&\\
weight=0.3812&weight=0.3857&\\\hline
-0.1208-0.2643i&-0.1490-0.2382i&0.9979\\
-0.1659-0.8150i&-0.1986-0.7942i&\\
0.4731&0.5009&\\\hline
\end{tabular}

\end{table}

This example shows a reasonably high accuracy of mixed state
reconstruction. The statistical properties of the proposed
algorithm were studied by means of the Monte-Carlo method. One
hundred numerical experiments were conducted similar to the one
described above. To verify the reliability, the solution was found
twice for each experiment (with random zero approximation
selection). The solutions appeared to be equal for all cases
(within negligible small computational error). The obtained
statistical fidelity distribution is shown in Fig.10. Numerical
research shows that the fidelity distribution density is well
described by the beta-distribution.

\begin{figure}
\includegraphics[width=0.5\textwidth]{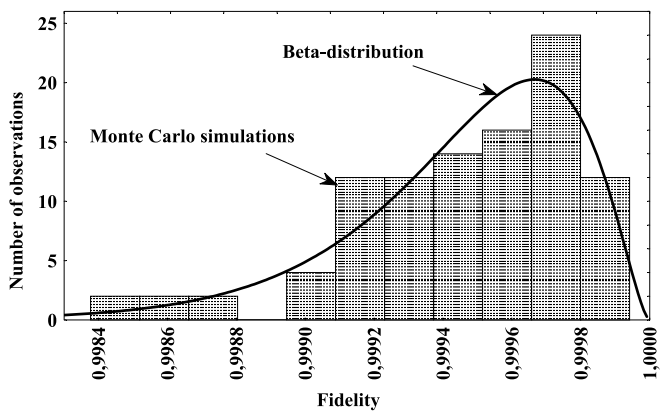}
\caption{Simulation of the fidelity between theoretical and
reconstructed density matrices in a mixture separation problem.
100 numerical experiments of 20,000 events per each (on the
average) were made.}
\end{figure}

\section{Conclusion}

The procedure of quantum state measurement for a three-state
optical system formed by a frequency- and spatially degenerate
two-photon field has been considered in this work. The method of
the statistical estimation of the quantum state through solving
the likelihood equation and examining the statistical properties
of the resulting estimates has been developed. Based on the
experimental data (fourth-order moments in the field) and the root
method of estimating quantum states, the initial wave function of
qutrits has been reconstructed.

Experimental data analysis is based on representing the event
generation intensity for each one of mutually-complementary
quantum processes as a squared module of some amplitude. A
complete set of measured processes amplitudes can be compactly
described using the instrumental matrix. In the framework of the
formalism of a process amplitude one can apply effective tools for
the quantum state reconstruction: least-squares and
maximum-likelihood methods.

The developed analysis tools provide means of quantum state reconstruction
from the experimental data with high accuracy and reliability. The estimate
accuracy is determined by concurrence of two types of errors: statistical
ones and instrumental ones. For smaller sample sizes statistical errors are
dominant, while for greater ones instrumental errors dominate.

Instrumental errors lead to fidelity saturation at less than unity level. In
the present work, fidelity for most of performed experiments (more than 20)
exceeded level of 0.995. For many cases the level of 0.9998 was achieved.

\begin{acknowledgments}
Useful discussions with A.Burlakov, A.Ekert, B.Englert,
D.Kazlikowski, and A.Lamas-Linares are gratefully acknowledged.

This work was supported in part by Russian Foundation of Basic
Research (projects 03-02-16444 and 02-02-16843) and the National
University of Singapore's Eastern Europe Research Scientist and
Student Programme. One of us (L.K.) acknowledges support from
INTAS-YS fellowship grant (Num. 03-55-1971).

\end{acknowledgments}

\appendix

\section{Statistical fluctuations of the state vector}

As it was already mentioned above, an un-normalized state vector
provides the most complete information about a quantum system. The
usage of un-normalized vector allows us to remove an interaction
constant in Eq. (24). The norm of the vector $c$, obtained as a
result of quantum system reconstruction, provides one with the
information about the total intensity of all the processes
considered in the experiment. However, the fluctuations of the
quantum state (and norm fluctuations, in particular) in a normally
functioning quantum information system should be within certain
range defined by the statistical theory. The present section is
devoted to this problem.

Practical significance of accounting for statistical fluctuations in a
quantum system relates to developing methods of estimation and control of
precision and stability of a quantum information system evolution, as well
as methods of detecting external interception (Eve's attack on the quantum
channel between Alice and Bob).

The estimate of the un-normalized state vector $c$, obtained by
the maximum-likelihood principle, differs from the exact state
vector $c^{(0)}$ by random values $\delta c = c^{(0)} - c$. Let us
consider the statistical properties of the fluctuation vector
$\delta c$ by expansion of the log likelihood function near the
stationary point,

\begin{equation}\delta \ln L = - \left[ \frac{1}{2}\left( {K_{sj} \delta c_s
\delta c_j + K_{sj}^\ast \delta c_s^\ast \delta c_j^\ast } \right)
+ I_{sj} \delta c_s^\ast \delta c_j  \right].\end{equation}
Together with the Hermitian matrix of the Fisher information $I$
(32), we define the symmetric Fisher information matrix $K$, whose
elements are defined by the following equation:
\begin{equation}K_{sj} = \sum\limits_\nu {\frac{k_\nu }{M_\nu ^2 }X_{\nu s}
X_{\nu j} } ,\end{equation} where $M_\nu $ is the amplitude of the
$\nu $- th process. In the general case, $K$ is a complex
symmetric non-Hermitian matrix. From all possible types of
fluctuations, let us pick out the so-called gauge fluctuations.
Infinitesimal global gauge transformations of a state vector are
as follows:

\begin{equation}\delta c_j = i\varepsilon c_j, j = 1,2,...,s\end{equation}
where $\varepsilon $ is an arbitrary small real number, $s$ is the
Hilbert space dimension.

Evidently, for gauge transformations $\delta \ln L = 0$. It means that two
state vectors that differ by a gauge transformation, are statistically
equivalent, i.e. they have the same likelihood. Such vectors are physically
equivalent since the global phase of the state vector is non-observable.
From statistical point of view, the set of mutually complementing
measurements should be chosen in such a way that for all other fluctuations
(except gauge fluctuations) $\delta \ln L < 0$. This inequality serves as
the statistical completeness condition for the set of mutually complementing
measurements.

Let us derive some constructive criteria of the statistical
completeness of measurements. The complex fluctuation vector
$\delta c$ is conveniently represented by a real vector of double
length. After extracting the real and the imaginary parts of the
fluctuation vector $\delta c_j = \delta c_j^{(1)} + i\delta
c_j^{(2)} $ we transfer from the complex vector $\delta c$ to the
real one $\delta \xi $:

\begin{equation}\delta c = \left( {{\begin{array}{*{20}c}
 {\delta c_1 } \hfill \\
 {:} \hfill \\
 {\delta c_s } \hfill \\
\end{array} }} \right) \to \delta \xi = \left( {{\begin{array}{*{20}c}
 {\delta c_1^{(1)} } \hfill \\
 : \hfill \\
 {\delta c_s^{(1)} } \hfill \\
 {\delta c_1^{(2)} } \hfill \\
 : \hfill \\
 {\delta c_s^{(2)} } \hfill \\
\end{array} }} \right).\end{equation}

In the particular case of qutrits ($s=3$) this transition provides
us with a 6-component real vector instead of a 3-component complex
vector.

In the new representation Eq. (A1) becomes:

\begin{equation}\delta \ln L = - H_{sj} \delta \xi _s \delta \xi _j = - \langle {\delta
\xi } |H| {\delta \xi } \rangle ,\end{equation} where matrix $H$
is the "complete information matrix" possessing the following
block form:

\begin{equation}H = \left( {{\begin{array}{*{20}c}
 {\textrm{Re}( {I + K} )} \hfill & { -\textrm{Im}( {I + K} )} \hfill \\
 {\textrm{Im}( {I - K} )} \hfill & {\textrm{Re}( {I - K} )} \hfill \\
\end{array} }} \right).\end{equation}

The matrix $H$ is real and symmetric. It is of double dimension
respectively to the matrices $I $ and $K$. For qutrits, $I$ and
$K$ are $3\times3$ matrices, while $H$ is $6\times6$.

Using matrix $H$ it is easy to formulate the desired characteristic
completeness condition for a mutually complementing set of measurements. For
a set of measurements to be statistically complete, it is necessary and
sufficient that one and only one eigenvalue of the complete information
matrix $H$ is equal to zero, while the other ones are strictly positive.

We would like to stress that checking the condition one not only verifies
the statistical completeness of a measurement protocol but also, insures
that the obtained extremum is of maximum likelihood.

An eigenvector that has eigenvalue equal to zero corresponds to gauge
fluctuation direction. Such fluctuations do not have physical meaning as
stated above. Eigenvectors corresponding to the other eigenvalues specify
the direction of fluctuations in the Hilbert space.

The principal fluctuation variance is

\begin{equation}\sigma _j^2 = \frac{1}{2h_j }, \quad j = 1,...2s - 1,\end{equation}
where $h_{j}$ is the eigenvalue of the information matrix $H$,
corresponding to the $j$-th principal direction.

The most critical direction in the Hilbert space is the one with
the maximum variance $\sigma _j^2 $, while the corresponding
eigenvalue $h_{j}$ is accordingly minimal. The knowledge of the
numerical dependence of statistical fluctuations allows one to
estimate distributions of various statistical characteristics.

The important information criterion that specifies the general
possible level of statistical fluctuations in quantum information
system is the chi-square criterion. It can be expressed as

\begin{equation}2\langle {\delta \xi }|H| {\delta \xi } \rangle
 \propto\chi ^2( {2s - 1} ),\end{equation}
where $s$ is the Hilbert space dimension

The left-hand side of Eq. (A8), which describes the level of state
vector information fluctuations, is a chi-square distribution with
$2s - 1$ degrees of freedom.

Validity of the analytical expression (A8) is justified by the
results of numerical modeling and observed data. Similarly to Eq.
(A4), let us introduce the transformation of a complex state
vector to a real vector of double length:

\begin{equation}c = \left( {{\begin{array}{*{20}c}
 {c_1 } \hfill \\
 {:} \hfill \\
 {c_s } \hfill \\
\end{array} }} \right) \to \xi = \left( {{\begin{array}{*{20}c}
 {c_1^{(1)} } \hfill \\
 : \hfill \\
 {c_s^{(1)} } \hfill \\
 {c_1^{(2)} } \hfill \\
 : \hfill \\
 {c_s^{(2)} } \hfill \\
\end{array} }} \right)\end{equation}
It can be shown that the information carried by a state vector is
equal to the doubled total number of observations in all
processes:

\begin{equation}\langle \xi |H| \xi \rangle = 2n,\end{equation}
where $n = \sum\limits_\nu {k_\nu } $.

Then, the chi-square criterion can be expressed in the form invariant to the
state vector scale (let us remind that we consider a non-normalized state
vector).

\begin{equation}\frac{\langle {\delta \xi } |H| {\delta \xi } \rangle
}{\langle \xi |H| \xi\rangle } \propto \frac{\chi ^2( {2s - 1}
)}{4n}\end{equation} Relation (A11) describes the distribution of
relative information fluctuations. It shows that relative
information uncertainty of a quantum state decreases with the
number of observations as $1/n$.

The mean value of relative information fluctuations is

\begin{equation}\frac{\overline {\langle {\delta \xi } |H| {\delta \xi }
\rangle } }{\langle \xi |H| \xi \rangle } = \frac{2s -
1}{4n}\end{equation}

The information fidelity may be introduced as a measure of
correspondence between the theoretical state vector and its
estimate:

\begin{equation}F_H = 1 - \frac{\langle {\delta \xi } |H| {\delta \xi }
\rangle }{\langle \xi |H| \xi \rangle } \quad .\end{equation}
Correspondingly, the value $1 - F_H $ is the information loss.

The convenience of $F_H $ relies on its simpler statistical properties
compared to the conventional fidelity $F$. For a system where statistical
fluctuations dominate, fidelity is a random value, based on the chi-square
distribution,

\begin{equation}F_H = 1 - \frac{\chi ^2( {2s - 1} )}{4n},\end{equation} where
$\chi ^2(2s - 1)$ is a random value of chi-square type with $2s -
1$ degrees of freedom.

Information fidelity asymptotically tends to unity when the sample size is
growing up. Complementary to statistical fluctuations noise leads to a
decrease in the informational fidelity level compared to the theoretical
level (A14).

\end{document}